\documentclass[twocolumn,english,conference,letterpaper]{IEEEtran}
\usepackage[T1]{fontenc}
\usepackage{geometry}
\usepackage{color}
\usepackage{units}
\usepackage{amsmath}
\usepackage{amsthm}
\usepackage{amssymb}
\usepackage{graphicx}

\makeatletter
\theoremstyle{plain}
\newtheorem{thm}{\protect\theoremname}

\usepackage{babel}
\usepackage[caption=false,font=footnotesize]{subfig}
\usepackage{balance}

\usepackage{tikz}

\usepackage{pgfplots}
\usepgfplotslibrary{groupplots}
\usepackage{graphicx}
\pgfplotsset{compat=newest}
\usepackage{units}
\usepackage{amsmath, amsbsy, amssymb}
\usepackage{tikzscale}
\usepackage{color}
\usetikzlibrary{plotmarks}
\usetikzlibrary{dsp,chains}
\usetikzlibrary{spy}
\usetikzlibrary{positioning,chains,fit,shapes,calc}
\pgfplotsset{
colormap={jetlight}{rgb = (  1.00000000,   1.00000000,   1.00000000),rgb = (  0.99607843,   0.99607843,   0.99816176),rgb = (  0.99215686,   0.99215686,   0.99644608),rgb = (  0.98823529,   0.98823529,   0.99485294),rgb = (  0.98431373,   0.98431373,   0.99338235),rgb = (  0.98039216,   0.98039216,   0.99203431),rgb = (  0.97647059,   0.97647059,   0.99080882),rgb = (  0.97254902,   0.97254902,   0.98970588),rgb = (  0.96862745,   0.96862745,   0.98872549),rgb = (  0.96470588,   0.96470588,   0.98786765),rgb = (  0.96078431,   0.96078431,   0.98713235),rgb = (  0.95686275,   0.95686275,   0.98651961),rgb = (  0.95294118,   0.95294118,   0.98602941),rgb = (  0.94901961,   0.94901961,   0.98566176),rgb = (  0.94509804,   0.94509804,   0.98541667),rgb = (  0.94117647,   0.94117647,   0.98529412),rgb = (  0.93725490,   0.93725490,   0.98529412),rgb = (  0.93333333,   0.93333333,   0.98541667),rgb = (  0.92941176,   0.92941176,   0.98566176),rgb = (  0.92549020,   0.92549020,   0.98602941),rgb = (  0.92156863,   0.92156863,   0.98651961),rgb = (  0.91764706,   0.91764706,   0.98713235),rgb = (  0.91372549,   0.91372549,   0.98786765),rgb = (  0.90980392,   0.90980392,   0.98872549),rgb = (  0.90588235,   0.90588235,   0.98970588),rgb = (  0.90196078,   0.90196078,   0.99080882),rgb = (  0.89803922,   0.89803922,   0.99203431),rgb = (  0.89411765,   0.89411765,   0.99338235),rgb = (  0.89019608,   0.89019608,   0.99485294),rgb = (  0.88627451,   0.88627451,   0.99644608),rgb = (  0.88235294,   0.88235294,   0.99816176),rgb = (  0.87843137,   0.87843137,   1.00000000),rgb = (  0.87450980,   0.87647059,   1.00000000),rgb = (  0.87058824,   0.87463235,   1.00000000),rgb = (  0.86666667,   0.87291667,   1.00000000),rgb = (  0.86274510,   0.87132353,   1.00000000),rgb = (  0.85882353,   0.86985294,   1.00000000),rgb = (  0.85490196,   0.86850490,   1.00000000),rgb = (  0.85098039,   0.86727941,   1.00000000),rgb = (  0.84705882,   0.86617647,   1.00000000),rgb = (  0.84313725,   0.86519608,   1.00000000),rgb = (  0.83921569,   0.86433824,   1.00000000),rgb = (  0.83529412,   0.86360294,   1.00000000),rgb = (  0.83137255,   0.86299020,   1.00000000),rgb = (  0.82745098,   0.86250000,   1.00000000),rgb = (  0.82352941,   0.86213235,   1.00000000),rgb = (  0.81960784,   0.86188725,   1.00000000),rgb = (  0.81568627,   0.86176471,   1.00000000),rgb = (  0.81176471,   0.86176471,   1.00000000),rgb = (  0.80784314,   0.86188725,   1.00000000),rgb = (  0.80392157,   0.86213235,   1.00000000),rgb = (  0.80000000,   0.86250000,   1.00000000),rgb = (  0.79607843,   0.86299020,   1.00000000),rgb = (  0.79215686,   0.86360294,   1.00000000),rgb = (  0.78823529,   0.86433824,   1.00000000),rgb = (  0.78431373,   0.86519608,   1.00000000),rgb = (  0.78039216,   0.86617647,   1.00000000),rgb = (  0.77647059,   0.86727941,   1.00000000),rgb = (  0.77254902,   0.86850490,   1.00000000),rgb = (  0.76862745,   0.86985294,   1.00000000),rgb = (  0.76470588,   0.87132353,   1.00000000),rgb = (  0.76078431,   0.87291667,   1.00000000),rgb = (  0.75686275,   0.87463235,   1.00000000),rgb = (  0.75294118,   0.87647059,   1.00000000),rgb = (  0.74901961,   0.87843137,   1.00000000),rgb = (  0.74509804,   0.88051471,   1.00000000),rgb = (  0.74117647,   0.88272059,   1.00000000),rgb = (  0.73725490,   0.88504902,   1.00000000),rgb = (  0.73333333,   0.88750000,   1.00000000),rgb = (  0.72941176,   0.89007353,   1.00000000),rgb = (  0.72549020,   0.89276961,   1.00000000),rgb = (  0.72156863,   0.89558824,   1.00000000),rgb = (  0.71764706,   0.89852941,   1.00000000),rgb = (  0.71372549,   0.90159314,   1.00000000),rgb = (  0.70980392,   0.90477941,   1.00000000),rgb = (  0.70588235,   0.90808824,   1.00000000),rgb = (  0.70196078,   0.91151961,   1.00000000),rgb = (  0.69803922,   0.91507353,   1.00000000),rgb = (  0.69411765,   0.91875000,   1.00000000),rgb = (  0.69019608,   0.92254902,   1.00000000),rgb = (  0.68627451,   0.92647059,   1.00000000),rgb = (  0.68235294,   0.93051471,   1.00000000),rgb = (  0.67843137,   0.93468137,   1.00000000),rgb = (  0.67450980,   0.93897059,   1.00000000),rgb = (  0.67058824,   0.94338235,   1.00000000),rgb = (  0.66666667,   0.94791667,   1.00000000),rgb = (  0.66274510,   0.95257353,   1.00000000),rgb = (  0.65882353,   0.95735294,   1.00000000),rgb = (  0.65490196,   0.96225490,   1.00000000),rgb = (  0.65098039,   0.96727941,   1.00000000),rgb = (  0.64705882,   0.97242647,   1.00000000),rgb = (  0.64313725,   0.97769608,   1.00000000),rgb = (  0.63921569,   0.98308824,   1.00000000),rgb = (  0.63529412,   0.98860294,   1.00000000),rgb = (  0.63137255,   0.99424020,   1.00000000),rgb = (  0.62745098,   1.00000000,   1.00000000),rgb = (  0.62941176,   1.00000000,   0.99411765),rgb = (  0.63149510,   1.00000000,   0.98811275),rgb = (  0.63370098,   1.00000000,   0.98198529),rgb = (  0.63602941,   1.00000000,   0.97573529),rgb = (  0.63848039,   1.00000000,   0.96936275),rgb = (  0.64105392,   1.00000000,   0.96286765),rgb = (  0.64375000,   1.00000000,   0.95625000),rgb = (  0.64656863,   1.00000000,   0.94950980),rgb = (  0.64950980,   1.00000000,   0.94264706),rgb = (  0.65257353,   1.00000000,   0.93566176),rgb = (  0.65575980,   1.00000000,   0.92855392),rgb = (  0.65906863,   1.00000000,   0.92132353),rgb = (  0.66250000,   1.00000000,   0.91397059),rgb = (  0.66605392,   1.00000000,   0.90649510),rgb = (  0.66973039,   1.00000000,   0.89889706),rgb = (  0.67352941,   1.00000000,   0.89117647),rgb = (  0.67745098,   1.00000000,   0.88333333),rgb = (  0.68149510,   1.00000000,   0.87536765),rgb = (  0.68566176,   1.00000000,   0.86727941),rgb = (  0.68995098,   1.00000000,   0.85906863),rgb = (  0.69436275,   1.00000000,   0.85073529),rgb = (  0.69889706,   1.00000000,   0.84227941),rgb = (  0.70355392,   1.00000000,   0.83370098),rgb = (  0.70833333,   1.00000000,   0.82500000),rgb = (  0.71323529,   1.00000000,   0.81617647),rgb = (  0.71825980,   1.00000000,   0.80723039),rgb = (  0.72340686,   1.00000000,   0.79816176),rgb = (  0.72867647,   1.00000000,   0.78897059),rgb = (  0.73406863,   1.00000000,   0.77965686),rgb = (  0.73958333,   1.00000000,   0.77022059),rgb = (  0.74522059,   1.00000000,   0.76066176),rgb = (  0.75098039,   1.00000000,   0.75098039),rgb = (  0.75686275,   1.00000000,   0.74117647),rgb = (  0.76286765,   1.00000000,   0.73125000),rgb = (  0.76899510,   1.00000000,   0.72120098),rgb = (  0.77524510,   1.00000000,   0.71102941),rgb = (  0.78161765,   1.00000000,   0.70073529),rgb = (  0.78811275,   1.00000000,   0.69031863),rgb = (  0.79473039,   1.00000000,   0.67977941),rgb = (  0.80147059,   1.00000000,   0.66911765),rgb = (  0.80833333,   1.00000000,   0.65833333),rgb = (  0.81531863,   1.00000000,   0.64742647),rgb = (  0.82242647,   1.00000000,   0.63639706),rgb = (  0.82965686,   1.00000000,   0.62524510),rgb = (  0.83700980,   1.00000000,   0.61397059),rgb = (  0.84448529,   1.00000000,   0.60257353),rgb = (  0.85208333,   1.00000000,   0.59105392),rgb = (  0.85980392,   1.00000000,   0.57941176),rgb = (  0.86764706,   1.00000000,   0.56764706),rgb = (  0.87561275,   1.00000000,   0.55575980),rgb = (  0.88370098,   1.00000000,   0.54375000),rgb = (  0.89191176,   1.00000000,   0.53161765),rgb = (  0.90024510,   1.00000000,   0.51936275),rgb = (  0.90870098,   1.00000000,   0.50698529),rgb = (  0.91727941,   1.00000000,   0.49448529),rgb = (  0.92598039,   1.00000000,   0.48186275),rgb = (  0.93480392,   1.00000000,   0.46911765),rgb = (  0.94375000,   1.00000000,   0.45625000),rgb = (  0.95281863,   1.00000000,   0.44325980),rgb = (  0.96200980,   1.00000000,   0.43014706),rgb = (  0.97132353,   1.00000000,   0.41691176),rgb = (  0.98075980,   1.00000000,   0.40355392),rgb = (  0.99031863,   1.00000000,   0.39007353),rgb = (  1.00000000,   1.00000000,   0.37647059),rgb = (  1.00000000,   0.99019608,   0.37254902),rgb = (  1.00000000,   0.98026961,   0.36862745),rgb = (  1.00000000,   0.97022059,   0.36470588),rgb = (  1.00000000,   0.96004902,   0.36078431),rgb = (  1.00000000,   0.94975490,   0.35686275),rgb = (  1.00000000,   0.93933824,   0.35294118),rgb = (  1.00000000,   0.92879902,   0.34901961),rgb = (  1.00000000,   0.91813725,   0.34509804),rgb = (  1.00000000,   0.90735294,   0.34117647),rgb = (  1.00000000,   0.89644608,   0.33725490),rgb = (  1.00000000,   0.88541667,   0.33333333),rgb = (  1.00000000,   0.87426471,   0.32941176),rgb = (  1.00000000,   0.86299020,   0.32549020),rgb = (  1.00000000,   0.85159314,   0.32156863),rgb = (  1.00000000,   0.84007353,   0.31764706),rgb = (  1.00000000,   0.82843137,   0.31372549),rgb = (  1.00000000,   0.81666667,   0.30980392),rgb = (  1.00000000,   0.80477941,   0.30588235),rgb = (  1.00000000,   0.79276961,   0.30196078),rgb = (  1.00000000,   0.78063725,   0.29803922),rgb = (  1.00000000,   0.76838235,   0.29411765),rgb = (  1.00000000,   0.75600490,   0.29019608),rgb = (  1.00000000,   0.74350490,   0.28627451),rgb = (  1.00000000,   0.73088235,   0.28235294),rgb = (  1.00000000,   0.71813725,   0.27843137),rgb = (  1.00000000,   0.70526961,   0.27450980),rgb = (  1.00000000,   0.69227941,   0.27058824),rgb = (  1.00000000,   0.67916667,   0.26666667),rgb = (  1.00000000,   0.66593137,   0.26274510),rgb = (  1.00000000,   0.65257353,   0.25882353),rgb = (  1.00000000,   0.63909314,   0.25490196),rgb = (  1.00000000,   0.62549020,   0.25098039),rgb = (  1.00000000,   0.61176471,   0.24705882),rgb = (  1.00000000,   0.59791667,   0.24313725),rgb = (  1.00000000,   0.58394608,   0.23921569),rgb = (  1.00000000,   0.56985294,   0.23529412),rgb = (  1.00000000,   0.55563725,   0.23137255),rgb = (  1.00000000,   0.54129902,   0.22745098),rgb = (  1.00000000,   0.52683824,   0.22352941),rgb = (  1.00000000,   0.51225490,   0.21960784),rgb = (  1.00000000,   0.49754902,   0.21568627),rgb = (  1.00000000,   0.48272059,   0.21176471),rgb = (  1.00000000,   0.46776961,   0.20784314),rgb = (  1.00000000,   0.45269608,   0.20392157),rgb = (  1.00000000,   0.43750000,   0.20000000),rgb = (  1.00000000,   0.42218137,   0.19607843),rgb = (  1.00000000,   0.40674020,   0.19215686),rgb = (  1.00000000,   0.39117647,   0.18823529),rgb = (  1.00000000,   0.37549020,   0.18431373),rgb = (  1.00000000,   0.35968137,   0.18039216),rgb = (  1.00000000,   0.34375000,   0.17647059),rgb = (  1.00000000,   0.32769608,   0.17254902),rgb = (  1.00000000,   0.31151961,   0.16862745),rgb = (  1.00000000,   0.29522059,   0.16470588),rgb = (  1.00000000,   0.27879902,   0.16078431),rgb = (  1.00000000,   0.26225490,   0.15686275),rgb = (  1.00000000,   0.24558824,   0.15294118),rgb = (  1.00000000,   0.22879902,   0.14901961),rgb = (  1.00000000,   0.21188725,   0.14509804),rgb = (  1.00000000,   0.19485294,   0.14117647),rgb = (  1.00000000,   0.17769608,   0.13725490),rgb = (  1.00000000,   0.16041667,   0.13333333),rgb = (  1.00000000,   0.14301471,   0.12941176),rgb = (  1.00000000,   0.12549020,   0.12549020),rgb = (  0.98627451,   0.12156863,   0.12156863),rgb = (  0.97242647,   0.11764706,   0.11764706),rgb = (  0.95845588,   0.11372549,   0.11372549),rgb = (  0.94436275,   0.10980392,   0.10980392),rgb = (  0.93014706,   0.10588235,   0.10588235),rgb = (  0.91580882,   0.10196078,   0.10196078),rgb = (  0.90134804,   0.09803922,   0.09803922),rgb = (  0.88676471,   0.09411765,   0.09411765),rgb = (  0.87205882,   0.09019608,   0.09019608),rgb = (  0.85723039,   0.08627451,   0.08627451),rgb = (  0.84227941,   0.08235294,   0.08235294),rgb = (  0.82720588,   0.07843137,   0.07843137),rgb = (  0.81200980,   0.07450980,   0.07450980),rgb = (  0.79669118,   0.07058824,   0.07058824),rgb = (  0.78125000,   0.06666667,   0.06666667),rgb = (  0.76568627,   0.06274510,   0.06274510),rgb = (  0.75000000,   0.05882353,   0.05882353),rgb = (  0.73419118,   0.05490196,   0.05490196),rgb = (  0.71825980,   0.05098039,   0.05098039),rgb = (  0.70220588,   0.04705882,   0.04705882),rgb = (  0.68602941,   0.04313725,   0.04313725),rgb = (  0.66973039,   0.03921569,   0.03921569),rgb = (  0.65330882,   0.03529412,   0.03529412),rgb = (  0.63676471,   0.03137255,   0.03137255),rgb = (  0.62009804,   0.02745098,   0.02745098),rgb = (  0.60330882,   0.02352941,   0.02352941),rgb = (  0.58639706,   0.01960784,   0.01960784),rgb = (  0.56936275,   0.01568627,   0.01568627),rgb = (  0.55220588,   0.01176471,   0.01176471),rgb = (  0.53492647,   0.00784314,   0.00784314),rgb = (  0.51752451,   0.00392157,   0.00392157),rgb = (  0.50000000,   0.00000000,   0.00000000)}
}

\addto\captionsenglish{}
\usepgfplotslibrary{external}
\usetikzlibrary{pgfplots.external}
\definecolor{mittelblau}{RGB}{0, 126, 198}
\definecolor{violettblau}{cmyk}{0.9, 0.6, 0, 0}
\definecolor{rot}{RGB}{238, 28 35}
\definecolor{apfelgruen}{RGB}{140, 198, 62}
\definecolor{gelb}{RGB}{1, 221, 0}
\definecolor{orange}{RGB}{244, 111, 33}
\definecolor{pink}{RGB}{237, 0, 140}
\definecolor{lila}{RGB}{128, 10, 145}
\definecolor{hellgrau}{RGB}{224, 224, 224}
\definecolor{mittelgrau}{RGB}{128, 128, 128}
\definecolor{dunkelgrau}{RGB}{80,80,80}
\definecolor{anthrazit}{RGB}{19, 31, 31}

\definecolor{myblue}{RGB}{80,80,160} 
\definecolor{mygreen}{RGB}{80,160,80}
\definecolor{myorgange}{RGB}{204,102,0}
\definecolor{lightblue}{RGB}{51,153,255}

\makeatother

\usepackage{babel}
\providecommand{\theoremname}{Theorem}

\begin{document}

\title{Achievable Rate Region for Iterative Multi-User Detection via Low-cost
Gaussian Approximation}

\author{\IEEEauthorblockN{Xiaojie~Wang\IEEEauthorrefmark{1},~Chulong Liang\IEEEauthorrefmark{2},~Li~Ping\IEEEauthorrefmark{2},~and~Stephan~ten~Brink\IEEEauthorrefmark{1}}\IEEEauthorblockA{\IEEEauthorrefmark{1}Institute of Telecommunications, Pfaffenwaldring
47, University of Stuttgart, 70569 Stuttgart, Germany\\
Email: \{wang, tenbrink\}@inue.uni-stuttgart.de\\
\IEEEauthorrefmark{2}Department of Electronic Engineering, City University
of Hong Kong, Hong Kong SAR, China\\
Email: \{eeliping, chuliang\}@cityu.edu.hk}}
\maketitle
\begin{abstract}
We establish a multi-user extrinsic information transfer (EXIT) chart
area theorem for the interleave-division multiple-access (IDMA) scheme,
a special form of superposition coding, in multiple access channels
(MACs). A low-cost multi-user detection (MUD) based on the Gaussian
approximation (GA) is assumed. The evolution of mean-square errors
(MSE) of the GA-based MUD during iterative processing is studied.
We show that the $K$-dimensional tuples formed by the MSEs of $K$
users constitute a conservative vector field. The achievable rate
is a potential function of this conservative field, so it is the integral
along any path in the field with value of the integral solely determined
by the two path terminals. Optimized codes can be found given the
integration paths in the MSE fields by matching EXIT type functions.
The above findings imply that i) low-cost GA-based MUD can provide
near capacity performance; ii) the sum-rate capacity (region) can
be achieved independently of the integration path in the MSE fields;
and iii) the integration path can be an extra degree of freedom for
code design.
\end{abstract}

\section{Introduction}

Theoretically, successive interference cancellation (SIC) together
with time-sharing or rate-splitting can achieve the entire capacity
region \cite{DTseBookFund}. SIC involves subtraction of successfully
detected signals. If practical forward error control (FEC) codes are
used, each subtraction incurs an extra overhead in terms of either
power or rate relative to an ideal capacity-achieving code \cite[Fig. 13.3]{YHuLPNOMAbook}.
Such overheads accumulate during SIC steps, moving it away from the
capacity limit particularly when the number of users is large. Also,
both time-sharing and rate-splitting involve segmenting a data frame
of a user into several sub-frames. The reduced sub-frame length implies
reduced coding gain for a practical turbo or low-density parity-check
(LDPC) type code, which further worsens the accumulation of losses.

Iterative detection can alleviate the loss accumulation problem using
soft cancellation instead of hard subtraction. A turbo or LDPC code
involving iterative detection can be optimized by matching the so-called
extrinsic information transfer (EXIT) functions of two local processors
\cite{tBAllerton01,RichardsonLDPCDesignTIT01}. In a single-user point-to-point
channel, such matching can offer near capacity performance, as shown
the area properties \cite{AKtBAreaThBEC04IT,BNMSEChart07IT}.

Interleave-division multiple-access (IDMA) is a low-cost transmission
scheme for MACs \cite{LipingIDMATWC06}. A Gaussian approximation
(GA) of cross-user interference is key to a low-cost IDMA detector
\cite{LiPingIDMACL04}. For comparison, consider a common \textit{a
posteriori} probability (APP) multi-user detector (MUD) \cite{LipingIDMATWC06}
and let $K$ be the number of users. The per-user complexity of a
GA-based MUD remains roughly the same for all $K$, while that of
an APP-based MUD is exponential in $K$.

A question naturally arises: At such low cost, what is the achievable
performance of IDMA under GA-based MUD? Some partial answers to this
question are available. It is shown that IDMA is capacity approaching
when all users see the same channel \cite{ChulongCLIDMASCMA}. It
is also known that the GA-based MUD can achieve some points in the
capacity region for multiple-input multiple-output (MIMO) MACs \cite{LLiuTSP18}.

This paper provides a comprehensive analysis of the achievable performance
of IDMA under GA-based MUD. We approach the problem based on multi-dimensional
curve matching of EXIT type functions. Let $v_{k}$ be the mean-square
error (MSE) (i.e., the variance) for the GA-based MUD for user $k$,
with $v_{k}=0$ indicating perfect decoding. Using the mutual information
(MI) and minimum MSE (MMSE) theorem \cite{GuoSVMMSEmIITI05,BNMSEChart07IT},
we show that the achievable sum-rate can be evaluated using a line
integral along a valid path in the $K$-dimensional vector field $\boldsymbol{v}=\left[v_{1},v_{2},\cdots,v_{K}\right]^{T}$.
Furthermore, the integral is path-independent and its value is solely
determined by the two terminations. The path independence property
greatly simplifies the code optimization problem. We gain some interesting
insights from the discussions in this paper.
\begin{itemize}
\item A low-cost GA-based MUD can provide near optimal performance. In particular,
it is capacity-achieving for Gaussian signaling. 
\item FEC codes optimized for single-user channels may not be good choices
for MACs. The FEC codes should be carefully designed to match the
GA-based MUD, which facilitates iterative detection. We will provide
examples for the related code design. 
\item A multi-user area theorem of EXIT chart is established for the code
design.
\item The sum-rate capacity is a potential function in the field formed
by $\boldsymbol{v}$, which leads to the path independence property.
\item All points of the MAC capacity region are achievable using only one
FEC code per user. This avoids the loss related to the frame segmentation
of SIC as aforementioned. 
\item The above results can be extended to MIMO MAC channels straightforwardly.
\end{itemize}
We will provide simulation results to show that properly designed
IDMA can approach the sum-rate MAC capacity for various decoding paths
in the MSE field within $\unit[1]{dB}$.

\section{Iterative IDMA Receiver}

Consider a general $K$-user MAC system, which is described by 
\begin{equation}
y={\displaystyle \sum_{i=1}^{K}}\sqrt{P_{i}}h_{i}x_{i}+n
\end{equation}
where $P_{i}$ denotes the received signal strength of the $i$th
user's signal, $h_{i}$ denotes the fading coefficients of the user,
$x_{i}$ is the $i$th transmit complex-valued signal and $n$ is
the additive (circularly symmetric complex) white Gaussian noise (AWGN)
with zero mean and unit variance, i.e., $\mathcal{CN}\left(0,\sigma^{2}=1\right)$.

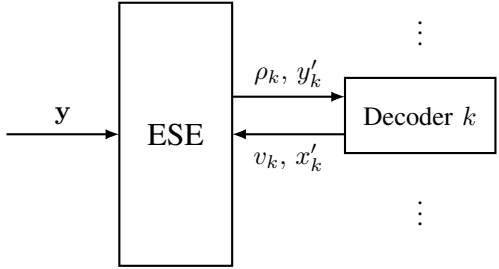
\begin{figure}[tbh]
\begin{centering}
\begin{tikzpicture} [>=latex]

%\draw [thick] (1,3) rectangle (3,2);
%\node at (2,2.5) {Decoder 1};
\draw [thick] (1,1) rectangle (3,0);
\node at (2,0.5) {Decoder $k$};
\node at (2,-0.7) {$\vdots$};
\node at (2, 1.7) {$\vdots$};
%\draw [thick] (1,-1.5) rectangle (3,-2.5);
%\node at (2,-2) {Decoder $K$};
\draw [thick] (-2,2) rectangle (-0.5,-1.5);
\node at (-1.25,0.25) {\large ESE};
\draw [thick,->](-3.5,0.25) -- (-2,0.25) node [midway, above] {$\mathbf{y}$};
%\draw [thick,->](-0.5,2.75) -- (1,2.75) node [midway, above] {$\rho_{1},\,y'_{1}$};
%\draw [thick,<-](-0.5,2.25) -- (1,2.25) node [midway, below] {$v_{1},\,x'_{1}$};
\draw [thick,->](-0.5,0.75) -- (1,0.75) node [midway, above] {$\rho_{k},\,y'_{k}$};
\draw [thick,<-](-0.5,0.25) -- (1,0.25) node [midway, below] {$v_{k},\,x'_{k}$};
%\draw [thick,->](-0.5,-1.75) -- (1,-1.75) node [midway, above] {$\rho_{K},\,y'_{K}$};
%\draw [thick,<-](-0.5,-2.25) -- (1,-2.25) node [midway, below] {$v_{K},\,x'_{K}$};
\end{tikzpicture}
\par\end{centering}
\caption{The iterative multi-user detection and decoding model.\label{fig:The-decoder-model-MU}}
\end{figure}

The iterative receiver is depicted in Fig. \ref{fig:The-decoder-model-MU}.
The elementary signal estimator (ESE) module has access to the channel
observation $y$ and feedbacks $x'_{k}$ from all the users' decoders.
It performs the so called soft interference cancellation (SoIC) and
provides each decoder a ``clean observation'', i.e., a signal with
reduced interference. The decoders perform its decoding based on the
signals $y'_{k}$ from ESE. 

To guarantee that the signals can be perfectly recovered, properly
designed channel codes shall be applied. This can be done by, e.g.,
EXIT chart based design (see \cite{AKtBAreaThBEC04IT}). 

\subsection{ESE functions}

The decoder feedbacks $x'_{k}$ are characterized by the MSE of its
estimates, denoted by $v_{i}$. The ``decoder observation'' after
SoIC is characterized by the signal-to-noise ratio (SNR), denoted
by $\rho_{i}$, assuming that the interference is Gaussian-distributed\footnote{The Gaussian assumption is valid for a large number of users with
arbitrary uncorrelated transmit symbols $x_{i}$ as the consequence
of the central limit theorem, or if the transmit signals $x_{i}$
are Gaussian by themselves.}, i.e.,
\begin{equation}
y'_{k}=\sqrt{\rho}_{k}x_{k}+z\tag{2}\label{eq:GaussAssum1}
\end{equation}
where $z$, comprised of AWGN and multi-user interference, is assumed
to be $\mathcal{CN}\left(0,1\right)$. This assumption greatly simplifies
the multi-user detection, henceforth it is referred to as GA-based
MUD. Therefore, the ESE transfer function for the $i$th user is given
by
\begin{equation}
\rho_{k}=\frac{P_{k}\left|h_{k}\right|^{2}}{{\displaystyle \sum_{j=1,j\ne i}^{K}}P_{j}\left|h_{j}\right|^{2}v_{j}+\sigma^{2}},\:\forall k=1,2,\ldots,K.\tag{3a}\label{eq:SINR}
\end{equation}
We can also express \eqref{eq:SINR} in a vector form as 
\begin{equation}
\boldsymbol{\rho}=\phi\left(\boldsymbol{v}\right)\tag{3b}\label{eq:SINRvec}
\end{equation}
where $\boldsymbol{\rho}=\left[\rho_{1},\rho_{2},\cdots,\rho_{K}\right]^{T}$
and $\boldsymbol{v}=\left[v_{1},v_{2},\cdots,v_{K}\right]^{T}$. Due
to the fact that the MSE is bounded by $0\leq v_{i}\leq\mathrm{E}\left[\left|x_{i}\right|^{2}\right]=1,$
we obtain that the SNR is also bounded by
\begin{equation}
\rho_{k,\mathrm{min}}=\frac{P_{k}\left|h_{k}\right|^{2}}{{\displaystyle \sum_{j=1,j\ne i}^{K}}P_{j}\left|h_{j}\right|^{2}+\sigma^{2}}\leq\rho_{k}\leq\frac{P_{k}\left|h_{k}\right|^{2}}{\sigma^{2}}=\rho_{k,\mathrm{max}}.\tag{{3c}}\label{eq:SINRRange}
\end{equation}
This bound implies that the single user decoder shall be able to decode
its signal before its ESE input reaches the maximum, i.e., $\exists\rho'_{k}\leq\rho_{k,\mathrm{max}},\:v_{k}\left(\rho'_{k}\right)=0.$

As the consequence of the iterative processing, we can write the SNR
and MSE vector as a function depending on a ``time'' or ``iteration''
variable $t$, i.e., $\boldsymbol{\rho}=\boldsymbol{\rho}\left(t\right),\,\boldsymbol{v}=\boldsymbol{v}\left(t\right)$
and $\boldsymbol{\rho}\left(t\right)=\phi\left(\boldsymbol{v}\left(t\right)\right).$

\subsection{DEC functions}

The decoder (DEC) transfer functions can be characterized by 
\begin{equation}
v_{k}=\psi_{k}\left(\rho_{k}\right),0\leq v_{k}\leq1,\:\forall k=1,2,\ldots,K.\tag{4a}\label{eq:DecFuncDef}
\end{equation}
Similar to \eqref{eq:SINRvec}, we can write \eqref{eq:DecFuncDef}
in a vector form as 
\begin{equation}
\boldsymbol{v}=\psi\left(\boldsymbol{\rho}\right).\tag{4b}\label{eq:DecFuncVec}
\end{equation}
Note that the boundaries for the uncooperative MAC\footnote{For MAC with cooperative encoders, a more general constraint can be
$\boldsymbol{1}^{T}\boldsymbol{v}\left(t=0\right)=K$ and $v_{k}\geqslant0,\:\forall k=1,2,\ldots,K$.} are given by 
\begin{align}
\boldsymbol{v} & =\psi\left(\boldsymbol{\rho}\left(t=0\right)\right)=\boldsymbol{1}\tag{4c}\label{eq:DecIni}\\
\boldsymbol{v} & =\psi\left(\boldsymbol{\rho}\left(t=\infty\right)\right)=\boldsymbol{0}\tag{4d}\label{eq:DecConv}
\end{align}
where $t$ is a variable which addresses the evolution of the SNR
$\boldsymbol{\rho}$ or MSE $\boldsymbol{v}$ through iterative processing.
\eqref{eq:DecIni} indicates that no a priori information is present
to the ESE at the beginning of iterations. \eqref{eq:DecConv} ensures
error-free decoding at the end. The decoders are typically APP decoders,
so that the MSEs $\boldsymbol{v}$ are also the conditional MMSE,
i.e., $v_{k}=\mathrm{E}\left[\left|x_{k}-\mathrm{E}\left[\left.x_{k}\right|x'_{k}\right]\right|^{2}\right]$.
Moreover, it is commonly assumed that the symbol estimates after APP
decoding can be modeled as an observation from the AWGN channel, i.e.,
\begin{equation}
x'_{k}=\sqrt{\rho'_{k}}x_{k}+w\tag{5}\label{eq:GaussAssum2}
\end{equation}
where $w$ follows $\mathcal{CN}\left(0,1\right)$. 

\subsection{Matching condition}

The matching codes which allow error-free decoding yet with highest
code rate (will be shown in Sec. III-A) shall satisfy
\begin{equation}
\psi\left(\boldsymbol{\rho}\left(t\right)\right)=\phi^{-1}\left(\boldsymbol{\rho}\left(t\right)\right).\tag{6}\label{eq:MatchCondLine}
\end{equation}
In other words, it is sufficient to match the code components along
a $K$-dimensional line, which is given by $\boldsymbol{\rho}\left(t\right)$.
It is noteworthy to mention that it is not necessary to match the
functions in the entire $K$-dimensional space, i.e., requiring $\psi\left(\boldsymbol{\rho}\right)=\phi^{-1}\left(\boldsymbol{\rho}\right),\,\forall\boldsymbol{\rho}$.
Matching along the path given by $\boldsymbol{\rho}\left(t\right)$
is much easier and achieves the MAC capacity (see Sec. III-A).

\section{Achievable rates}

The achievable rates under the GAs in \eqref{eq:GaussAssum1} and
\eqref{eq:GaussAssum2} can be written as \cite{GuoSVMMSEmIITI05,BNMSEChart07IT,XYuanITI14}
\begin{equation}
R_{k}={\displaystyle \int_{0}^{\infty}}f\left(\rho_{k}+f^{-1}\left(v_{k}\right)\right)d\rho_{k},\:\forall k=1,2,\ldots,K.\tag{7a}\label{eq:UserRateIntGeneral}
\end{equation}
where $f\left(\rho_{k}\right)=v_{k}$ denotes the MMSE. 

\subsection{Gaussian alphabets}

We consider Gaussian signals, i.e., $x_{i}$ are Gaussian-distributed
which can be achieved by using, e.g., superposition coded modulation
(SCM) \cite{LPSCMJSAC09}. Therefore, the MMSE is given by $f_{G}\left(\rho\right)=\frac{1}{1+\rho},$
and the achievable rates are 
\begin{multline}
\begin{split}R_{k}={\displaystyle \int_{0}^{\infty}}\frac{1}{\rho_{k}+v_{k}^{-1}}d\rho_{k} & =-{\displaystyle \int}_{v_{k}=1}^{v_{k}=0}\frac{g_{k}}{\mathbf{g}^{T}\boldsymbol{v}+\sigma^{2}}dv_{k},\\
\forall k=1,2,\ldots,K
\end{split}
\tag{7b}\label{eq:UserRateGaussian}
\end{multline}
where $\mathbf{g}^{T}=\left[P_{1}\left|h_{1}\right|^{2},P_{2}\left|h_{2}\right|^{2},\cdots,P_{K}\left|h_{K}\right|^{2}\right]^{T}$
contains the powers of all users, $\boldsymbol{v}=\left[v_{1},v_{2},\cdots,v_{K}\right]^{T}$
and $g_{k}=P_{k}\left|h_{k}\right|^{2}$ denotes the $k$th element
of vector $\mathbf{g}$. The derivation is shown in Appendix \ref{sec:Proof-of}.

The sum-rate of all users is 
\begin{align}
R_{\mathrm{sum}} & ={\displaystyle \sum_{k=1}^{K}}R_{k}=-{\displaystyle \int}_{L\left(t\right)}\frac{\mathbf{g}}{\mathbf{g}^{T}\boldsymbol{v}\left(t\right)+\sigma^{2}}\cdot d\boldsymbol{v}\left(t\right)\tag{8a}\label{eq:LineIntRsum-1}
\end{align}
where \eqref{eq:LineIntRsum-1} denotes a line integral defined by
$L=\boldsymbol{v}\left(t\right),\:t\in\left[0,\infty\right]$. It
can be further shown that the integrands constitute a gradient of
a scalar field (a.k.a., potential function), i.e., $\nabla_{\boldsymbol{v}}\mathrm{log}\left(\sigma^{2}+\mathbf{g}^{T}\boldsymbol{v}\right)=\frac{\mathbf{g}}{\mathbf{g}^{T}\boldsymbol{v}+\sigma^{2}}.$
Thus, the achievable sum-rate can be written as 
\begin{align*}
R_{\mathrm{sum}} & =-{\displaystyle \int_{L=\boldsymbol{v}\left(t\right)}}\left[\nabla\mathrm{log}\left(\sigma^{2}+\mathbf{g}^{T}\boldsymbol{v}\right)\right]\boldsymbol{v}'\left(t\right)dt\\
 & \overset{\eqref{eq:DecIni},\eqref{eq:DecConv}}{=}\mathrm{log}\left(1+\frac{\sum_{k=1}^{K}P_{k}\left|h_{k}\right|^{2}}{\sigma^{2}}\right)\tag{8b}
\end{align*}
which is independent of the path taken for code matching. In other
words, any path with matched DEC functions can achieve the sum-rate
capacity. The matching condition given in \eqref{eq:MatchCondLine}
is thus also proved, since it can be easily verified that $R_{k}<-{\displaystyle \int}_{v_{k}=1}^{v_{k}=0}\frac{g_{k}}{\mathbf{g}^{T}\boldsymbol{v}+\sigma^{2}}dv_{k}$
and thus $R_{\mathrm{sum}}<\mathrm{log}\left(1+\frac{\sum_{k=1}^{K}P_{k}\left|h_{k}\right|^{2}}{\sigma^{2}}\right)$,
if $\psi\left(\boldsymbol{\rho}\left(t\right)\right)<\phi^{-1}\left(\boldsymbol{\rho}\left(t\right)\right)$.
On the contrary, if $\psi\left(\boldsymbol{\rho}\left(t\right)\right)>\phi^{-1}\left(\boldsymbol{\rho}\left(t\right)\right)$,
error-free decoding is not possible.

This leads to the following theorem. 

\textit{\label{Assumptions:}Assumptions}\footnote{These assumptions have been widely used for turbo-type iterative receivers.
It is generally accepted that these assumptions are sufficiently accurate
for practical systems.}\textit{:}
\begin{enumerate}
\item Exchanged messages of the \textit{extrinsic} and \textit{a priori}
channel are observations from AWGN channels, given in \eqref{eq:GaussAssum1}
and \eqref{eq:GaussAssum2}.
\item The channel decoder satisfies the matching condition in \eqref{eq:MatchCondLine}
and has MAP (i.e., APP) performance.
\end{enumerate}
\begin{thm}
\label{thm:IDMA-is-capacity-achieving} Under the above assumptions,
the achievable sum-rate in IDMA with GA-based MUD for any path $L\left(t\right):\boldsymbol{v}_{s}=\mathbf{1}\rightarrow\boldsymbol{v}_{e}=\mathbf{0}$
(starting from $\boldsymbol{v}_{s}=\mathbf{1}$ to $\boldsymbol{v}_{e}=\mathbf{0}$)
is given by 
\begin{align*}
R_{\mathrm{sum}} & =-{\displaystyle \int}_{L\left(t\right)}f_{G}\left(\boldsymbol{\rho}\left(t\right)+f_{G}^{-1}\left(\boldsymbol{v}\left(t\right)\right)\right)\cdot d\boldsymbol{\rho}\left(t\right)\\
 & =\mathrm{log}\left(1+\frac{\sum_{k=1}^{K}P_{k}\left|h_{k}\right|^{2}}{\sigma^{2}}\right).
\end{align*}
\end{thm}
\begin{IEEEproof}
see above.
\end{IEEEproof}

\subsection{Finite alphabets}

If the symbols $x_{i}\in\mathcal{S}_{i}$ are taken from finite alphabets
$\left|\mathcal{S}_{i}\right|<\infty$, the capacity formula, in general,
can not be expressed in closed-form. However, eq. \eqref{eq:UserRateIntGeneral}
is still valid and can be used to evaluate, by numerical integrals,
the achievable rates. 

We show in \cite[Fig. 2]{MUAreaTBD} that the loss to Gaussian capacity
due to finite modulation can be approached by imposing a larger number
of users or data layers, depending on the target sum-rate. There,
we also provide numerical results showing that near-capacity performance
can be achieved with quadrature phase shift keying (QPSK). Due to
space limitation, we refer interested readers to\textit{\textcolor{magenta}{{}
}}\cite[Sec. III-B]{MUAreaTBD} for further discussions.

\subsection{Example: path vs rate tuples}

\begin{figure}[tbh]
\begin{centering}
\includegraphics[width=1\columnwidth]{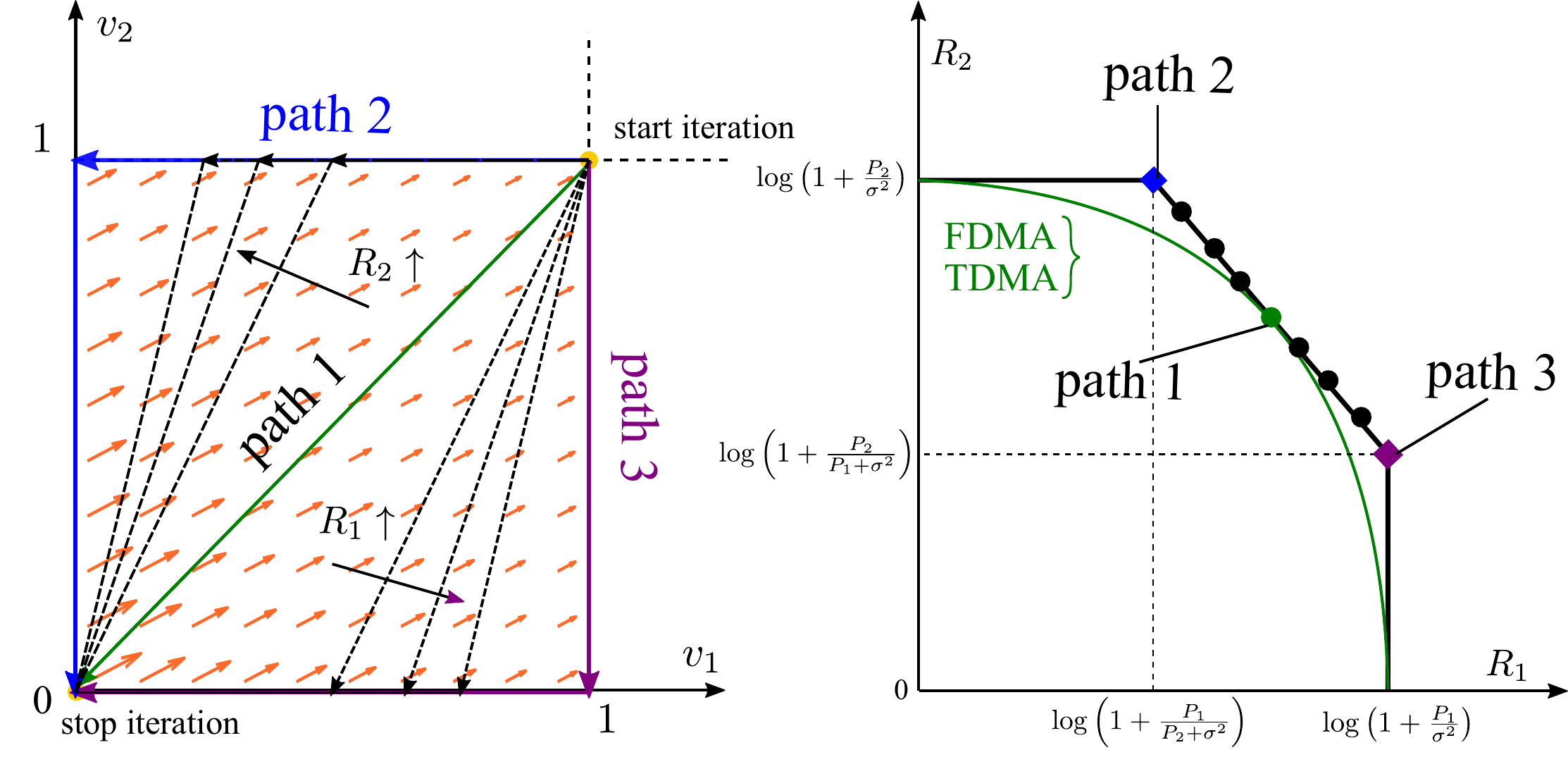}
\par\end{centering}
\caption{Illustration (exemplary for two users) of different integration paths
achieving different rate pairs $\left(R_{1},R_{2}\right)$; the arrows
in the left figure illustrate the two-dimensional MSE vector field;
the achieved rate pairs are marked in the right figure for the corresponding
paths.\label{fig:Illustration-of-different}}
\end{figure}

Consider a simple two-user case, i.e., $K=2$. Fig. \ref{fig:Illustration-of-different}
illustrates some special paths and their corresponding achievable
rate pairs. 
The simplest path is a straight line between the starting point $\boldsymbol{v}\left(t=0\right)=\mathbf{1}$
and the stop point $\boldsymbol{v}\left(t=\infty\right)=\mathbf{0}$,
denoted by \textit{path 1}. It is straightforward to obtain $R_{k}=\frac{g_{k}}{\mathbf{g}^{T}\mathbf{1}}\mathrm{log}\left(1+\frac{\mathbf{g}^{T}\mathbf{1}}{\sigma^{2}}\right),\,\forall k.$
In this case, the achievable rate of each user is proportional to
the received signal power strength $g_{k}$. For the two-user case,
the rate tuple coincides with the point where TDMA/FDMA achieves the
sum-rate capacity. In \textit{path 1, }it satisfies $v_{i}\left(t\right)=v_{j}\left(t\right),\,\forall i,j,\,t.$
The matching code for the $k$th user shall have the following MSE
characteristic function 
\[
v_{k}=\begin{cases}
1 & \rho_{k}\leq\rho_{k,\mathrm{min}}\\
\frac{1}{\mathbf{g}^{T}\mathbf{1}-g_{k}}\cdot\left(\frac{1}{\rho_{k}}-\sigma^{2}\right) & \rho_{k,\mathrm{min}}\leq\rho_{k}\leq\rho_{k,\mathrm{max}}\\
0 & \rho_{k}\geq\rho_{k,\mathrm{max}}
\end{cases}.
\]
\textit{Path 2 }and\textit{ path 3} are comprised of $K$ segments
and each segment satisfies $\frac{dv_{l}}{dt}\ne0$ and $\frac{dv_{k}}{dt}=0,\forall k\ne l$.
There exist $K!$ such paths, which constitute the $K!$ SIC points.
The decoding functions are step functions with sharp transitions
at threshold SNRs $\rho_{k,\mathrm{SIC}}$. This type of decoding
functions may pose difficulties for practical code designs, compared
to that with smooth transitions.

\subsection{Achievable rate region}

To achieve other points in the MAC capacity region, i.e., with maximum
sum-rate but different individual user rates, other paths shall be
used. In the following theorem, we show that the entire MAC capacity
region can be achieved by showing the existence of paths. Examples
for constructing a dedicated path achieving a feasible rate tuple
are provided in \cite[Sec. V-A, case 2]{MUAreaTBD}.
\begin{thm}
\label{thm:IDMA-achieves-everywhere}IDMA with GA-based MUD and the
assumptions in Theorem \ref{thm:IDMA-is-capacity-achieving} achieves
every rate tuple in the $K$-user MAC capacity region $\mathcal{C\left(\mathit{K}\right)}$.
Given a feasible target rate tuple $\mathbf{R}=\left[R_{1},R_{2},\cdots,R_{K}\right]\in\mathcal{C\left(\mathit{K}\right)}$,
there exists at least one path defined by $L\left(t\right)=\boldsymbol{v}\left(t\right):\boldsymbol{v}_{s}=\mathbf{1}\rightarrow\boldsymbol{v}_{e}=\mathbf{0}$
which achieves $\mathbf{R}$.
\end{thm}
\begin{IEEEproof}
See Appendix B.
\end{IEEEproof}
\textit{Remark: }It is easy to prove that there exists a unique path
for each of the $K!$ SIC corner points and the decoding functions
shall be step functions. For other rate tuples, it can be verified
that there exist many different paths achieving that rate tuple. The
choice of the integration path poses varying degrees of difficulty
for the design of matching codes. Thus, the design of an appropriate
integration path could be an extra degree of freedom for code design.

\textit{Numerical Results: }We designed matching (binary) LDPC codes
for $K=3$ with unequal-power distribution and QPSK for different
integration paths and rate tuples at sum-rate one. Bit error rate
(BER) simulations and density evolution results show that the gap
to Gaussian capacity is below $\unit[1]{dB}$ for all cases \cite[Sec. V]{MUAreaTBD}. 

\section{MU-MIMO Channel}

Assume that each transmitter has $N_{t,i}$ antennas and the receiver
has $N_{R}$ antennas respectively; then, the received signal can
be written as 
\begin{equation}
\mathbf{y}={\displaystyle \sum_{k=1}^{K}}\sqrt{P_{k}}\mathbf{H}_{k}\mathbf{x}_{k}+\mathbf{n}\tag{9}\label{eq:MIMOsysM}
\end{equation}
where $\mathbf{H}_{k}$ is the channel of the $k$th user, $\mathbf{n}$
denotes the uncorrelated noise $\mathrm{E}\left[\mathbf{n}\mathbf{n}^{H}\right]=\sigma^{2}\mathbf{I}$.
In this case, the ESE module is replaced by a linear MMSE (LMMSE)
receiver \cite{XYuanITI14}. Under the LMMSE-based ESE, the SNR of
user $k$ can be written as \cite{YuanMIMOLMMSESINR}
\begin{equation}
\rho_{k}=\frac{{\displaystyle \sum_{i=1}^{N_{t,k}}}\mathbf{h}_{k,i}^{H}\mathbf{R}^{-1}\mathbf{h}_{k,i}}{1-{\displaystyle v_{k}\sum_{i=1}^{N_{t,k}}}\mathbf{h}_{k,i}^{H}\mathbf{R}^{-1}\mathbf{h}_{k,i}}\tag{10}
\end{equation}
where $\mathbf{h}_{k,i}$ denotes the $i$th column of the $k$th
user's channel matrix $\mathbf{H}_{k}$ and 
\[
\mathbf{R}=\sigma^{2}\mathbf{I}+\mathbf{HVH}^{H}
\]
with $\mathbf{V}=\mathrm{diag}\left(P_{1}v_{1},P_{2}v_{2},\cdots,P_{K}v_{K}\right)$
and $\mathbf{H}$ being the concatenated channels of all users. Following
a similar approach in Appendix A,  the sum-rate can be obtained as
\begin{align}
R_{\mathrm{sum}} & ={\displaystyle \sum_{i=1}^{K}R_{i}}=-\int_{\mathbf{v}=\mathbf{1}}^{\mathbf{v}=\mathbf{0}}\nabla\mathrm{log\,det}\left[\mathbf{R}\right]d\mathbf{v}\nonumber \\
 & =\mathrm{log\,det}\left[\mathbf{I}+\frac{1}{\sigma_{n}^{2}}\mathbf{H}^{H}\mathbf{P}\mathbf{H}\right]\tag{11}
\end{align}
where $\mathbf{P}=\mathrm{diag}\left(P_{1},P_{2},\cdots,P_{K}\right)$.
Path independence follows from the condition 
\[
\frac{\partial}{\partial v_{k}}\mathrm{log}\mathrm{\,det}\left[\mathbf{R}\right]=\mathrm{trace}\left[\mathbf{R}^{-1}\mathbf{H}_{k}\mathbf{H}_{k}^{H}\right]={\displaystyle \sum_{i=1}^{N_{t,k}}}\mathbf{h}_{k,i}^{H}\mathbf{R}^{-1}\mathbf{h}_{k,i}
\]
with Jacobi's formula.

\section{Conclusion}

It is proved that the simple interleave-division multiple-access (IDMA),
relying on a low-cost Gaussian approximation (GA) based multi-user
detector (MUD), is capacity-achieving for general Gaussian multiple
access channels (GMAC) with arbitrary number of users, power distribution
and with single or multiple antennas. We show that IDMA with matching
codes is capacity-achieving for arbitrary decoding path in the mean-square
error (MSE) vector field. This property is further used to prove that
IDMA achieves not only the sum-rate capacity, but the entire GMAC
capacity region. The construction of capacity-achieving codes is also
provided by establishing the area theorem for multi-user extrinsic
information transfer (EXIT) chart.

\appendices{}

\section{\label{sec:Proof-of}Proof of \eqref{eq:UserRateGaussian}}

Let $\rho_{k,\mathrm{max}}$ and $\rho_{k,\mathrm{min}}$ as defined
in \eqref{eq:SINRRange}. The achievable rates can be thus expressed
as

\begin{align*}
R_{k} & ={\displaystyle \int_{\rho_{k,\mathrm{min}}}^{\rho_{k,\mathrm{max}}}}\frac{1}{\rho_{k}+v_{k}^{-1}}d\rho_{k}+{\displaystyle \int_{0}^{\rho_{k,\mathrm{min}}}}\frac{1}{\rho_{k}+1}d\rho_{k}\\
 & \overset{\rho'_{k}=\frac{d\rho_{k}}{dv_{k}}}{=}{\displaystyle \int_{v_{k}=1}^{v_{k}=0}}\frac{\rho'_{k}}{\rho_{k}+v_{k}^{-1}}dv_{k}+{\displaystyle \int_{0}^{\rho_{k,\mathrm{min}}}}\frac{1}{\rho_{k}+1}d\rho_{k}\\
 & ={\displaystyle \int_{1}^{0}}\frac{\rho'_{k}-v_{k}^{-2}+v_{k}^{-2}}{\rho_{k}+v_{k}^{-1}}dv_{k}+\underset{=w_{0}}{\underbrace{\mathrm{log}\left(1+\rho_{k,\mathrm{min}}\right)}}\\
 & =\left[\mathrm{log}\left(\rho_{k}+v_{k}^{-1}\right)+{\displaystyle \int_{1}^{0}}\frac{v_{k}^{-2}}{\rho_{k}+v_{k}^{-1}}dv_{k}\right]_{v_{k}=1}^{^{v_{k}=0}}+w_{0}\\
\overset{(1\mathrm{a})}{=} & \left[\mathrm{log}\left(\rho_{k}+v_{k}^{-1}\right)+{\displaystyle \int}\left(v_{k}^{-1}-\frac{g_{k}}{\mathbf{g}^{T}\boldsymbol{v}+\sigma^{2}}\right)dv_{k}\right]_{v_{k}=1}^{^{v_{k}=0}}\\
 & +w_{0}\\
 & =\left[\mathrm{log}\left(\rho_{k}v_{k}+1\right)-{\displaystyle \int}\frac{g_{k}}{\mathbf{g}^{T}\boldsymbol{v}+\sigma^{2}}dv_{k}\right]_{v_{k}=1}^{^{v_{k}=0}}+w_{0}\\
 & =-{\displaystyle \int}_{1}^{0}\frac{g_{k}}{\mathbf{g}^{T}\boldsymbol{v}+\sigma^{2}}dv_{k}
\end{align*}
where $g_{k}=P_{k}\left|h_{k}\right|^{2}$ is the $k$th element of
the vector $\mathbf{g}$.

\section{Proof of Theorem \ref{thm:IDMA-achieves-everywhere}}

The user rate $R_{k}=-{\displaystyle \int}_{v_{k}=1}^{v_{k}=0}\frac{g_{k}}{\mathbf{g}^{T}\boldsymbol{v}+\sigma^{2}}dv_{k}$
is obviously a continuous and monotone decreasing function of $\boldsymbol{v}$.
If $v_{k}$ are unbounded, then $R_{k}$ are unbounded with the single
sum-rate constraint ${\displaystyle \sum R_{k}}\leq\mathrm{log}\left(\frac{\mathbf{g}^{T}\boldsymbol{1}+\sigma^{2}}{\sigma^{2}}\right)$.
However, the value range of $R_{k}$ is constrained by the fact that
$0\leq v_{l}\leq1,\forall l$. Therefore, it is bounded by 
\[
R_{k}\leq-{\displaystyle \int}_{v_{k}=1}^{v_{k}=0}\frac{g_{k}}{g_{k}v_{k}+\sigma^{2}}dv_{k}=\mathrm{log}\left(\frac{g_{k}+\sigma^{2}}{\sigma^{2}}\right)
\]
and similarly $R_{k}\geq\mathrm{log}\left(\frac{g_{k}+\sigma^{2}}{\sum_{l\ne k}g_{l}+\sigma^{2}}\right)$.
Further, the constraints on $v_{l}$ leads to 
\begin{align*}
R_{k}+R_{l} & \leq\mathrm{log}\left(\frac{g_{k}+g_{l}+\sigma^{2}}{\sigma^{2}}\right),\forall k\ne l\\
R_{k}+R_{l}+R_{m} & \leq\mathrm{log}\left(\frac{g_{k}+g_{l}+g_{m}+\sigma^{2}}{\sigma^{2}}\right),\forall k\ne l\ne m\\
\vdots & \vdots\\
{\displaystyle \sum R_{k}} & \leq\mathrm{log}\left(\frac{\mathbf{g}^{T}\boldsymbol{1}+\sigma^{2}}{\sigma^{2}}\right)
\end{align*}
and these constraints constitute the capacity region.

\balance

\bibliographystyle{IEEEtran}
\bibliography{bibliography}

\end{document}